\documentclass[journal=jctcce,manuscript=article]{achemso}
\setkeys{acs}{maxauthors=0,articletitle=true}
\usepackage{achemso}
\usepackage{placeins}
\usepackage{graphics}
\usepackage{amssymb,amsfonts}
\usepackage{graphicx}
\usepackage[table,dvipsnames]{xcolor}
\usepackage{multirow}
\usepackage{caption}
\usepackage{subcaption}
\usepackage{booktabs}
\usepackage{colortbl}
\usepackage{amsmath}
\usepackage{amsopn}
\usepackage{bm}
\usepackage{braket}
\usepackage{siunitx}
\usepackage{color}
\usepackage{array}
\usepackage{lscape}
\usepackage{mciteplus}
\usepackage[version=3]{mhchem}
\usepackage{ulem}
\usepackage{listings}
\usepackage{enumerate}
\usepackage{lmodern}
\usepackage{mhchem}
\usepackage[implicit=false]{hyperref}
\usepackage[T1]{fontenc}

\author{Frederik {\O}. Kjeldal}
\affiliation{DTU Chemistry, Technical University of Denmark\\Kemitorvet Bldg. 206, 2800 Kgs. Lyngby, Denmark}
\author{Janus J. Eriksen}
\email{janus@dtu.dk}
\affiliation{DTU Chemistry, Technical University of Denmark\\Kemitorvet Bldg. 206, 2800 Kgs. Lyngby, Denmark}

\title[TITLE]{Transferability of Atom-Based Neural Networks}

\begin{document}

\begin{abstract}

Machine-learning models in chemistry---when based on descriptors of atoms embedded within molecules---face essential challenges in transferring the quality of predictions of local electronic structures and their associated properties across chemical compound space. In the present work, we make use of adversarial validation to elucidate certain intrinsic complications related to machine inferences of unseen chemistry. On this basis, we employ invariant and equivariant neural networks---both trained either exclusively on total molecular energies or a combination of these and data from atomic partitioning schemes---to evaluate how such models scale performance-wise between datasets of fundamentally different functionality and composition. We find the inference of local electronic properties to improve significantly when training models on augmented data that appropriately expose local functional features. However, molecular datasets for training purposes must themselves be sufficiently comprehensive and rich in composition to warrant any generalizations to larger systems, and even then, transferability can still only genuinely manifest if the body of atomic energies available for training purposes exposes the uniqueness of different functional moieties within molecules. We demonstrate this point by comparing machine models trained on atomic partitioning schemes based on the spatial locality of either native atomic or molecular orbitals.

\end{abstract}

\newpage

\section{Introduction}

The success of supervised machine learning (ML) in applications to electronic-structure problems relies fundamentally on the ability of such models to transfer performance in predictions from known to unknown chemistry. Given how ML generally excels at interpolation rather than extrapolation to out-of-distribution data, significant challenges are necessarily faced when seeking to infer chemical properties for molecular datasets that are fundamentally different to those available for training and validation. The enhancement of extrapolation performance is therefore not merely a question of designing sufficiently flexible and physically motivated model architectures but also of curating necessary diversity within the underlying training pool, an either costly or inherently scarce resource in the overall design process.\\

One prominent form of such transferability is that which rules across {\textit{conformational space}}, e.g., when using ML as a vehicle for simulations of molecular dynamics where the number of atoms and chemical composition are kept fixed. Here, unseen chemistry explored in various regions on a potential energy surface must resemble that used to train the ML model in a way that will facilitate predictions to predominantly rely on interpolations. This has been a popular application of ML in quantum chemistry ever since machine potentials first came into existence~\cite{blank1995neural, gassner1998representation,lorenz2004representing,behler2007generalized,handley2009optimal,handley2010potential,behler2011neural}, and it continues to be as much in vogue today as never before~\cite{li2015molecular,hansen2015machine,botu2015adaptive,rupp2015machine,chmiela2018towards,christensen2020role,unke2021machine,behler2021four}.\\

An arguably more difficult challenge lies in the application of supervised models to molecular problems of arbitrary composition, especially given how training pools are typically limited in scope~\cite{rupp2012fast,smith2017ani,tkatchenko_mueller_ml_qc_jcp_2018}. The task of learning chemistry across {\textit{compositional space}} in a transferable manner is thus contingent on the ability to generalize inferences from smaller to larger and, possibly, more complex molecular systems. Not only will the design of sophisticated encodings of unique atomic environments matter, but so will also the abundance of diverse functional motifs in the training data. The ruling premise here is that if local chemical environments are to be learned and generalized, a spatial localization of the chemistry at hand must somehow be enforced. In the course of the present work, we will demonstrate how to accomplish exactly this by exposing intrinsically local features within the training data, features which are indeed transferable. Qualified predictions of local energy contributions have potential use in genetic algorithms~\cite{hammer_ml_qc_jctc_2018,meldgaard2018machine,jensen_chem_space_chem_sci_2019}, and changes to these along reaction coordinates can help elucidate key chemical concepts, such as, selectivity, reactivity, and stability~\cite{kjeldal2023localprop,schutt2017quantum,unke2018reactive}.\\

Over the years, a wealth of different ML architectures have been proposed. The arguably most successful of these are the so-called high-dimensional neural networks (HDNNs), in which the total energy of a molecule is decomposed into a sum of $N$ atomic contributions,
\begin{align}
    E = \sum_i^N \mathcal{E}_i \ .
    \label{eq:hdnn}
\end{align}
This decomposition inherently allows the model to scale to arbitrary size by treating each atom locally. Descriptors of the local chemical environments around atoms are encoded somehow, and a feed-forward neural network is used to calculate an energy for each atom based on these. Originally, HDNNs were designed to model high-dimensional potential energy surfaces (hence their name), but the architecture has since been used to transfer and generalize property predictions across limited regions of chemical compound space as well.\\

Two main flavors of HDNNs exist, differing from one another primarily in how the local chemical environment of an atom embedded within a molecule is represented and, implicitly, how it interacts with its neighbours. Traditional HDNNs are based on fixed analytical descriptors of local chemical environments in terms of atom-centered symmetry functions that encode two- and three-body information within a certain spatial region~\cite{behler2007generalized}. Graph-based message-passing neural networks (MPNNs) instead encode such local environments by iteratively exchanging geometric information between atoms through convolutions over neighbors, a process which principally allows for the inclusion of long-range interactions~\cite{gilmer2017neural,unke2019physnet,schutt2017schnet}.. Most recently, MPNNs relying on equivariances rather than just invariances have started emerging, whereby more (angular) information gets encoded into the representation of atoms~\cite{batzner20223,musaelian2023learning}.\\

While the architecture of neural network models depend crucially on locality assumptions, only modest attention has been paid to the resulting atomic energies or the effect of including reference values for these in the training data. In the present work, we propose the use of local chemical information, namely decomposed atomic energies calculated from electronic-structure methods, as a means to improve the transferability of ML models. By reformulating the loss function from a global to a local quantity, we interpolate between different local chemical moieties instead of directly extrapolating to unseen chemistry, and by incorporating local atomic energies directly in the loss function, we constrain the minimization of the total energy error. This guides the optimization of the neural networks toward other minima, thus permitting different generalization properties. In particular, for low-data tasks, the inclusion of atomic energies steers the network to yield more physically sound atomic energies over those returned by a network trained exclusively on total energies.\\

The present study is outlined as follows. In Sect. \ref{comp_detail_sect}, we present the specific atomic decomposition schemes, datasets, and machine-learning models used throughout. Sect. \ref{res_sect} covers both proof-of-concept and more realistic experiments relating to transferability across functional and compositional space, while Sect. \ref{concl_sect} provides some conclusions and an outlook.

\section{Computational Details}
\label{comp_detail_sect}

In Sect. \ref{decomp_sect}, we begin by introducing the two types of atomic partitioning schemes that we will study within the present work. Next, our MPNN architecture of choice and the training protocol are discussed in Sect. \ref{nn_train_sect}, before we provide details on our different datasets alongside a brief introduction to the adversarial validation of these in Sects. \ref{dataset_sect} and \ref{adv_val_theory_sect}.

\subsection{Atomic Decomposition Schemes}\label{decomp_sect}

As discussed in a recent study of ours~\cite{kjeldal2023localprop}, the total molecular energy of a given system at the level of Kohn-Sham density functional theory (KS-DFT) may be decomposed amongst its atoms based on the spatial locality of either its atomic (AOs) or molecular orbitals (MOs). Specifically, in the standard energy density analysis (EDA) scheme of Nakai~\cite{nakai_eda_partitioning_cpl_2002,nakai_eda_partitioning_ijqc_2009}, one partitions the full 1-electron reduced density matrix (1-RDM) on account of which atoms individual AOs are localized on, achieved by simply limiting all necessary trace operations in the energy functional to only those basis functions that are spatially assigned to individual atoms. In the MO-based scheme of Eriksen~\cite{eriksen_decodense_jcp_2020}, on the other hand, atom-specific 1-RDMs are constructed via a set of 1-RDMs unique to the individual occupied MOs and a set of appropriate weights that distribute these among all constituent atoms. These are then the principal 1-RDM objects used to evaluate the KS-DFT energy functional. While the AO-based EDA decomposition is invariant with respect to orbital rotations, a suitable combination of localized MOs and corresponding populations is required in the MO-based analogue. As has previously been demonstrated~\cite{eriksen_elec_ex_decomp_jcp_2022,eriksen_local_condensed_phase_jpcl_2021,kjeldal2023decomposing,zamok2024decomposing}, intrinsic bond orbitals (IBOs) and Mulliken-like population weights determined in an intermediate basis of intrinsic atomic orbitals (IAOs) constitute excellent choices~\cite{knizia_iao_ibo_jctc_2013,lehtola_jonsson_pm_jctc_2014}, owing to their stability upon a change of AO basis and ease of chemical interpretation. All decompostions have been performed in the {\texttt{decodense}} code~\cite{decodense}.\\

In contrast to the electronic-structure decomposition schemes discussed above, an atomic partitioning may also be inferred from vast amounts of quantum-chemical data. For instance, in the application of HDNNs, one naturally obtains quantities popularly referred to as atomic energies from the chemical locality assumption underpinning Eq.~\ref{eq:hdnn}. These energies essentially serve as additional degrees of freedom that allow for the NN architecture to scale to systems of different composition and size; earlier studies have sought to investigate the physical relevance of data-derived atomic energies, e.g., in the stability of aromatic rings or for use in evolutionary algorithms~\cite{schutt2017quantum, hammer_ml_qc_jctc_2018,unke2018reactive}. Be that as it may, one obvious drawback of these decompositions is the sensitive dependency on the underlying data, which may negatively conflate local chemical information and, in turn, prevent the transferability of atomic properties.

\subsection{Neural Network Architecture and Training}\label{nn_train_sect}

Throughout the present study, we will use {\texttt{NequIP}}~\cite{batzner20223}---a leading equivariant MPNN---for training all of our proposed machine models~\bibnote{For the model configuration, we used the standard {\texttt{NequIP}} defaults. Example configuration files can be found among the SI as YAML files. In general, no optimization of hyperparameters was done to ensure a fair comparison between the different models.}. Our loss function is of mean square error (MSE) type with separate weightings of total ($E$) and atomic ($\mathcal{E}$) energy error contributions,
\begin{align}
    \mathcal{L} = \frac{1}{N}\big[\lambda_E \sum_i^N(\hat{E}_i - E_i )^2 + \lambda_{\mathcal{E}} \sum_i^N\sum_k^{N^{(i)}_{\mathrm{atoms}}} (\hat{\mathcal{E}}_{i,k} - \mathcal{E}_{i,k})^2 \big] \ . \label{loss_funct_eq}
\end{align}

To train a data-driven decomposition scheme, only the total energies of a given dataset matter, that is, we set $\lambda_{\mathcal{E}} \equiv 0$ (denoting these as {\textit{total energy}} models). For the models trained also on atomic energies from electronic-structure decompositions (EDA or IBO/IAO), we use a uniform weighting, $\lambda_{E} \equiv \lambda_{\mathcal{E}} \equiv 1$. The inclusion of total energies has previously been found to be important, as these regularize errors in atomic energies to cancel more favorably~\cite{kjeldal2023decomposing}. In training our networks, 80\% of a given dataset is used for training, leaving 20\% for validation, and the test set is trivially kept separate and used only after the network has been trained. Throughout our study, atomic energies will be reported as contributions to molecular atomization energies, that is, with respect to isolated atoms in the gas phase.

\subsection{Datasets}\label{dataset_sect}

To illustrate some of the inherent difficulties in transferring predictions across different kinds of chemistry, we have curated a number of small datasets with exclusive chemical motifs, namely, hydroxyls, carbonyls, as well as primary and secondary amines. For instance, the exercise of predicting atomization energies of carbonyl-containing compounds by means of a model trained exclusively on molecules containing hydroxyl functional groups is deliberately unrealistic; but, as we will discuss, it may provide key insights into more realistic chemical problems of transferability. These modest datasets have all been derived from QM7~\cite{blum_gdb13_2009}.\\

\begin{table}[ht!]
    \centering
    \begin{tabular}{c|c|c|c|c}
        Name & Atomic composition & Heavy atoms & Size & Parent dataset  \\
        \hline 
        Hydroxyl & H, C, O & $3-7$ & 389 & QM7 \\
        Carbonyl & H, C, O & $3-7$ & 283 & QM7 \\
        \hline 
        Primary amine & H, C, N & $3-7$ & 389 & QM7 \\
        Secondary amine & H, C, N & $3-7$ & 405 & QM7 \\
        \hline 
        QM7 & H, C, N, O, S & $1-7$ & 7,165 & GDB13 \\
        QM13$^*$ & H, C, N, O, S & 13 & 3,553 & GDB13 \\
        \hline 
        QM9~\bibnote{The pruned QM9 dataset of Ref. \citenum{ramakrishnan2014quantum} contained a total of 130,831 entries. From the single-point calculations in {\texttt{PySCF}} at the B3LYP/pcseg-1 level of theory performed in the course of the present study, alongside the subsequent orbital localizations needed in {\texttt{decodense}}, a total of 125,761 molecules passed all convergence and stability checks.} & H, C, N, O, F & $1-9$ & 125,761 & GDB17 \\
        QM17$^*$ & H, C, N, O, F & 17 & 4,670 & GDB17 \\
    \end{tabular}
    \caption{Key information about the datasets used in the present study.}
    \label{tab:datasets}
\end{table}
Next, two larger datasets have been designed to probe transferability in transitioning from small to larger and more complex molecular systems. The so-called QM13$^*$ and QM17$^*$ datasets are derived from the parent GDB13 and GDB17 datasets~\cite{blum_gdb13_2009,ruddigkeit_gdb17_2012}, respectively, by retaining only entries that consist of exactly 13 and 17 non-hydrogen atoms. In the case of  QM13$^*$, 5,000 random molecules built from H, C, N, O, and S atoms were extracted from the GDB13 dataset so as to align with the chemical composition of QM7. The geometry of each of these molecules was optimized at the B3LYP/6-31G(2df,p) level of theory in {\texttt{Gaussian16}}~\cite{becke_b3lyp_functional_jcp_1993,frisch_b3lyp_functional_jpc_1994,frisch2016gaussian}. Single-point calculations in {\texttt{PySCF}}~\cite{pyscf_wires_2018,pyscf_jcp_2020} and energy decompositions in {\texttt{decodense}}~\cite{decodense} were subequently performed at the B3LYP/pcseg-1 level of theory~\cite{jensen_pc_basis_sets_jcp_2001}, resulting in 3,553 entries of the dataset~\bibnote{From the geometry optimizations, 3,814 molecules converged with no imaginary frequencies, of which 3,553 successfully passed a subsequent SCF stability check in {\texttt{PySCF}}.}. Likewise, QM17$^*$ consists of 4,670 entries drawn from a random pool of 10k molecules from GDB17, of which the QM9 dataset is also a subset~\cite{ramakrishnan2014quantum}. Table \ref{tab:datasets} provides detailed information on the composition of all datasets of the present study.

\subsection{Adversarial Validation}\label{adv_val_theory_sect}

Beyond statistics about chemical composition, such as, molecular size, atom types, functional motifs, etc., tangible differences between datasets, particularly on a single-molecule level, may still prove difficult to quantify. One commonly used approach for comparing two chemical datasets involves calculating descriptors for each molecule before performing some form of unsupervised learning, e.g., clustering or dimensionality reductions~\cite{butina1999unsupervised,schneider2015development} The computation of these descriptors is computationally inexpensive, but the procedure has the distinct disadvantage of relying on fixed chemical descriptors, descriptors which, in the case of MPNNs, are iteratively learned rather than precomputed. One therefore runs the risk of losing or capturing fundamentally different chemical trends than the NN architecture used to train the actual energy regression model in question. As an alternative, one can make use of the latent space (or internal representation) of a trained machine model as the designated description vector prior to applying a subsequent unsupervised clustering algorithm~\cite{shrivastava2021fragnet,routh2021latent}.\\

As yet another option, so-called adversarial validation can be used to gauge differences between datasets so as to explain and predict where a given ML model may be expected to suffer inference errors. In adversarial validation~\cite{adv_val_2016,adv_val_2023}, two or more datasets are combined and shuffled, upon which a classifier is trained to untangle the datasets, predicting a net label for each molecule. If the classifier is trivially able to discern which dataset a given molecule originally belongs to, the two datasets are categorized as being sufficiently dissimilar in their local chemistry (and {\textit{vice versa}} if the classification is less successful). Since adversarial validation makes use of the original datasets as reference values, this allows for an educated guess at how well the performance of a trained model will transfer to a new dataset before even running any simulations. Furthermore, the validation will rely on the same model architecture as the energy regression and may thus allow for a fine-grained examination of a given dataset. This is particularly fitting for our purposes herein, given how our models have atom-level resolution that allow for the inspection of single atoms or functional groups.\\

One technical note on this type of analysis is warranted. In order to fairly evaluate the classification of any two datasets, a relatively even balance must exist between the volume (composition) of these. Except for the QM9/QM17$^*$ pair, this happens to be the case and we therefore generally included all data points in our adversarial validation (80 \% retained for training and validation with the remaining 20 \% reserved for evaluating the performance of the classifier). The large discrepancy between the QM9 and QM17$^*$ datasets, however, demanded some reductions to the former. As such, a random subset of QM9 was selected to limit the combined number of molecules to 10k in this specific adversarial validation.

\section{Results}\label{res_sect}

\subsection{Adversarial Validation}\label{adv_val_res_sect}

Starting with our proof-of-concept comparison of the four functional datasets discussed in Sect. \ref{dataset_sect}, one would trivially expect the classifier to be able to detect differences at the molecular level, but perhaps less so between the atoms of any two datasets due to many near-identical scaffolds. As evidenced by the accuracy of the classifier in Table \ref{tab:adversarial_validation}, alcohols are perfectly distinguished from aldehydes and ketones and primary from secondary amines. The same is observed to hold true for the individual oxygens and nitrogens of the datasets, thus verifying that the model correctly identifies the functional groups containing these as the single most important part of the classification of the four different kinds of molecules.\\

\begin{figure}[ht!]
    \centering
    \includegraphics[width=0.6\textwidth]{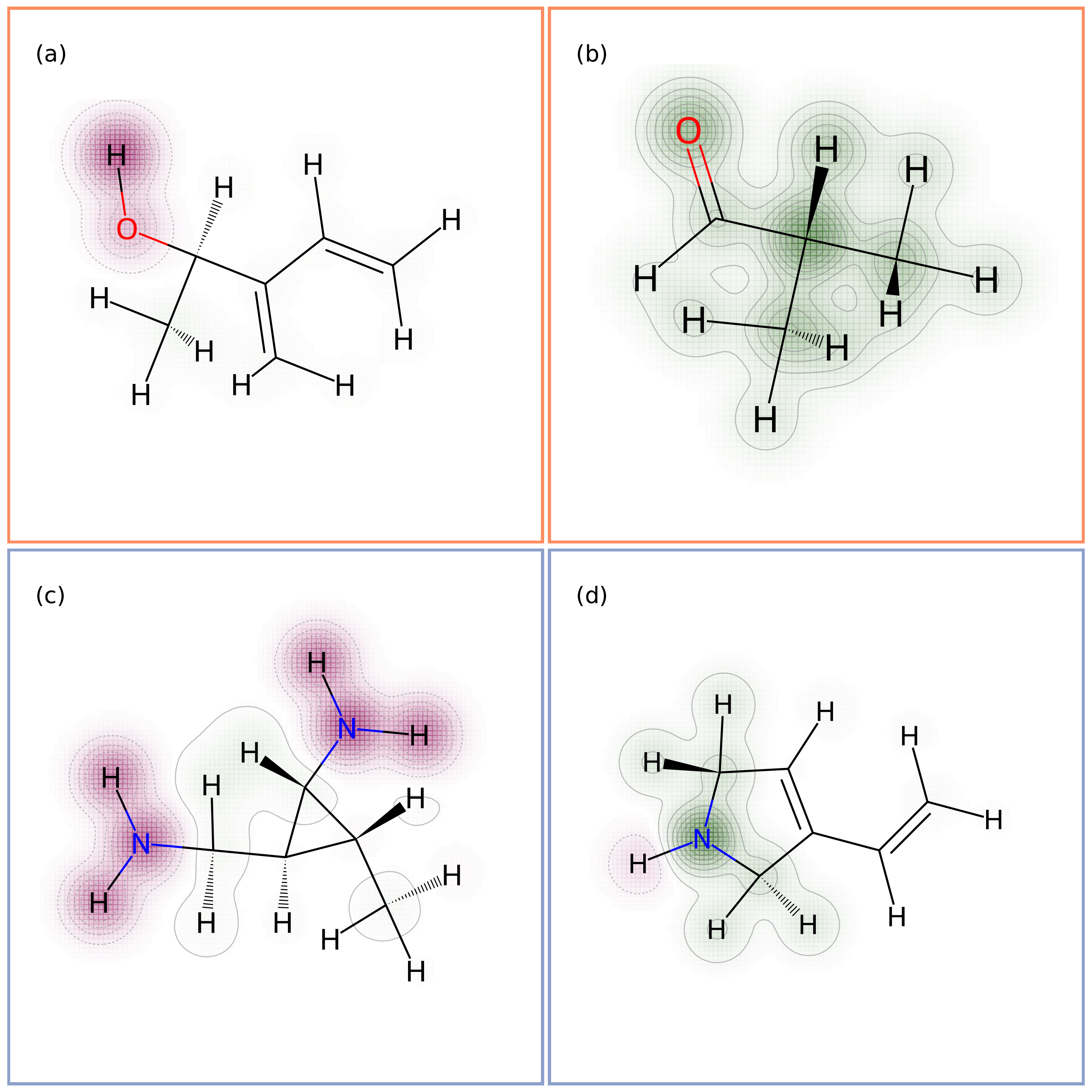}
    \caption{Contour plots of the classifier logits (cf. text for details) for random examples of $(a)$ hydroxyls, $(b)$ carbonyls, $(c)$ primary amines, and $(d)$ secondary amines. Red and green contours correspond to negative and positive logits, respectively, which indicate predictions of hydroxyls and primary amines (red) and carbonyls and secondary amines (green).}
    \label{toy_grid_fig}
\end{figure}

\begin{table}[ht!]
    \centering
    \begin{tabular}{c||c|c|c|c|c|c|c|c}
         Dataset & Mol. & Atom. & H & C & N & O & S & F \\
         \hline
         Hydroxyl/Carbonyl & 1.00 & 0.51 & 0.43 & 0.53 & -- & 1.00 & -- & -- \\
         Prim./Sec. Amines & 1.00 & 0.62 & 0.62 & 0.55 & 1.00 & -- & -- & -- \\
         \hline\hline
         QM7/QM13$^*$ & 1.00 & 0.72 & 0.61 & 0.90 & 0.83 & 0.70 & 0.80 & -- \\
         QM9/QM17$^*$ & 0.98 & 0.65 & 0.54 & 0.79 & 0.80 & 0.74 & -- & 0.87 \\
    \end{tabular}
    \caption{Accuracy, $x$, of the adversarial validation classifier across the different datasets for molecules (Mol.), atoms (Atom.), and individual elements (H, C, N, O, S, and F). Accuracies of $x = 0$ and $x = 1$ indicate complete failure or success in the classification, respectively.}
    \label{tab:adversarial_validation}
\end{table}
However, many of the hydrogens and carbons are clearly misclassified. Using the logits of each atom classification (i.e., the atomic outputs from the neural network before applying a sigmoid activation function), whenever these have large amplitudes it will imply that the model has a high confidence in attributing an atom. By plotting these amplitudes as contours superimposed on molecular 2D structures, we can visually identify the important parts of a (mis)classification for a given molecule. Results of this type are presented in Fig. \ref{toy_grid_fig}.\\

In Fig. \ref{toy_grid_fig}, we observe how the classification of the alcohol in question focuses almost exclusively on both constituents of the hydroxyl group, with the hydrogen seemingly the most important atom. For the classification of the aldehyde, on the other hand, the foci of the model are less evident. While the carbonyl oxygen is obviously important to the overall classification, it is much less integral than in the hydroxyl case, with all the atoms in the molecule contributing to the correct classification. In the case of the two amine datasets, the nitrogens are similarly always correctly classified (cf. Table \ref{tab:adversarial_validation}), while all other atoms are harder to distinguish between the primary and secondary amines. In fact, some of the hydrogens, either attached to or adjacent to a nitrogen, are even misclassified, which appears to indicate that the local electronic structures associated with these two functional groups are more alike than for hydroxyls and carbonyls. This is arguably to be expected on the basis of chemical intuition alone. We here reiterate how these four datasets are intentionally pathological in that they are limited in scope, and inferences of electronic structures present in one dataset from those in another should be unfeasible. For instance, the atomic energies of heterocyclic amines are known to differ significantly from those of primary ones~\cite{kjeldal2023localprop}, and an ML model should thus have no sound basis for predicting the former if trained only on the latter.\\

\begin{figure}[ht!]
    \centering
    \includegraphics[width=0.6\textwidth]{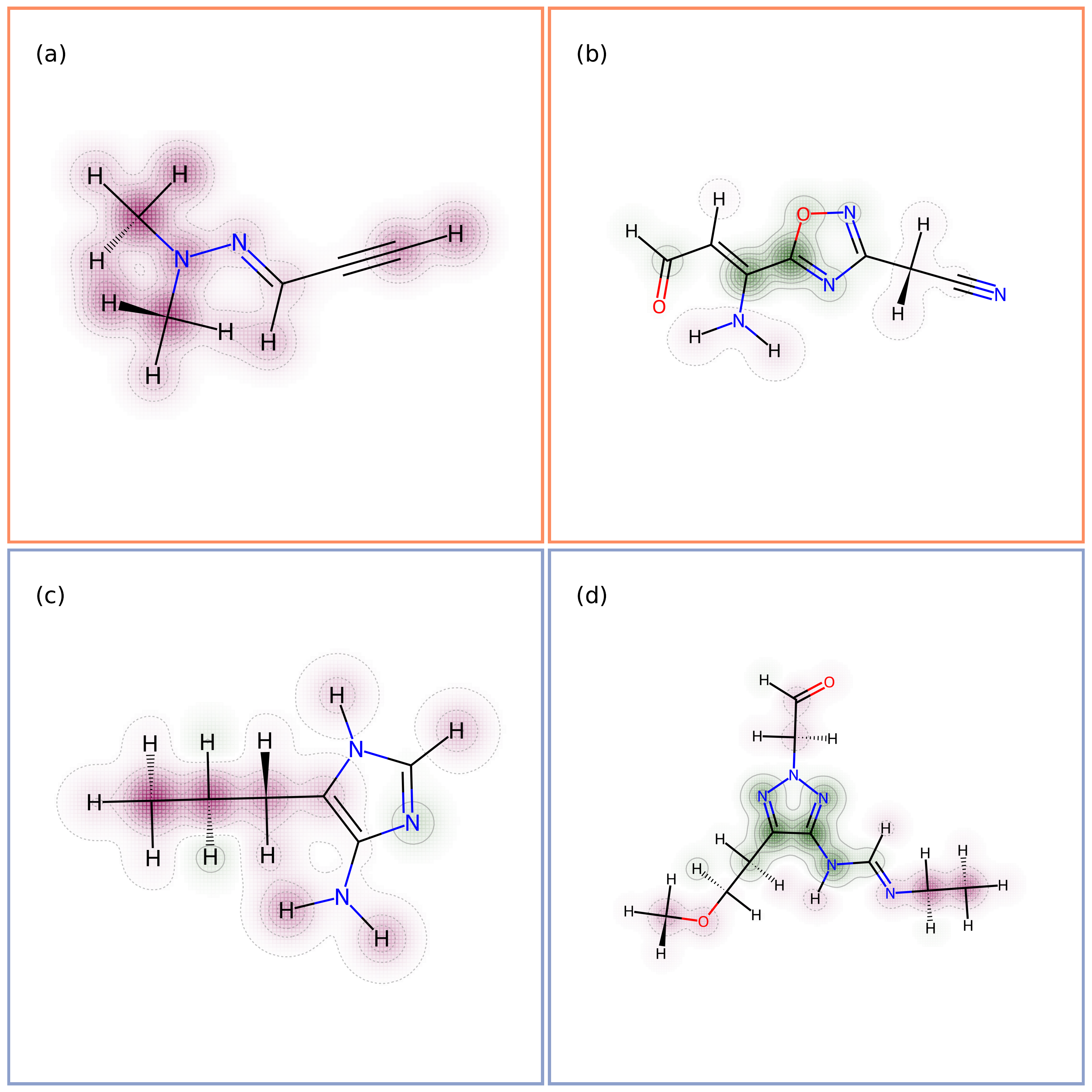}
    \caption{Contour plots of the classifier logits for random entries of the $(a)$ QM7, $(b)$ QM13$^*$, $(c)$ QM9, and $(d)$ QM17$^*$ datasets. As in Fig. \ref{toy_grid_fig}, red/green contours correspond to negative/positive logits, indicating predictions of QM7/9 (red) and QM13$^*$/17$^*$ (green).}
    \label{qmx_grid_fig}
\end{figure}
For the QM7/9 and QM13$^*$/17$^*$ datasets in Table \ref{tab:adversarial_validation} instead, given how carbon atoms constitute the main backbone of all the molecules of any of these, a fair assumption would be that, upon letting molecules grow in size, the most central carbons should be readily classified as belonging to any of the QM13$^*$/17$^*$ datasets. On the other hand, those atoms that reside near or within terminal groups should be harder to classify. An exception to this rule is fluorine, which has a seemingly high accuracy in the adversarial validation across the QM9/QM17$^*$ datasets, despite being a terminal atom. This is a result of the low number of molecules that contain F atoms in this restricted analysis (cf. Sect. \ref{adv_val_theory_sect}); 23 of these belong to QM9 and only 4 to QM17$^*$, which makes the classifier predict all fluorines as belonging to the former (as statistically evidenced by a Matthews correlation coefficient of 0.0).\\

From the random examples in Fig. \ref{qmx_grid_fig}, we generally find these trends to align well with the predicted logits. While the limited sizes of the molecules of QM7 and---to some extent---QM9 allow for positive distinctions from those of QM13$^*$ and QM17$^*$, respectively, it is predominantly the innermost elements that are successfully classified as belonging to the larger datasets. As an implication, training an ML model on QM7/9 and next applying it to QM13$^*$/17$^*$ should yield small errors for peripheral atoms, with increasing atomic errors upon moving towards the center of the larger molecules of 13 and 17 heavy atoms each, respectively. Also, given how limited in composition the QM7 dataset is (7k molecules, as opposed to more than 125k in QM9), one would expect to see significantly larger overall errors for a model trained on QM7 and applied to QM13$^*$ than one trained on QM9 and applied to QM17$^*$. In the following, both of these conjectures will be numerically asserted.

\subsection{Functional Transferability}\label{func_trans_res_sect} 

\begin{figure}[htb]
    \centering
    \includegraphics[width=0.7\textwidth]{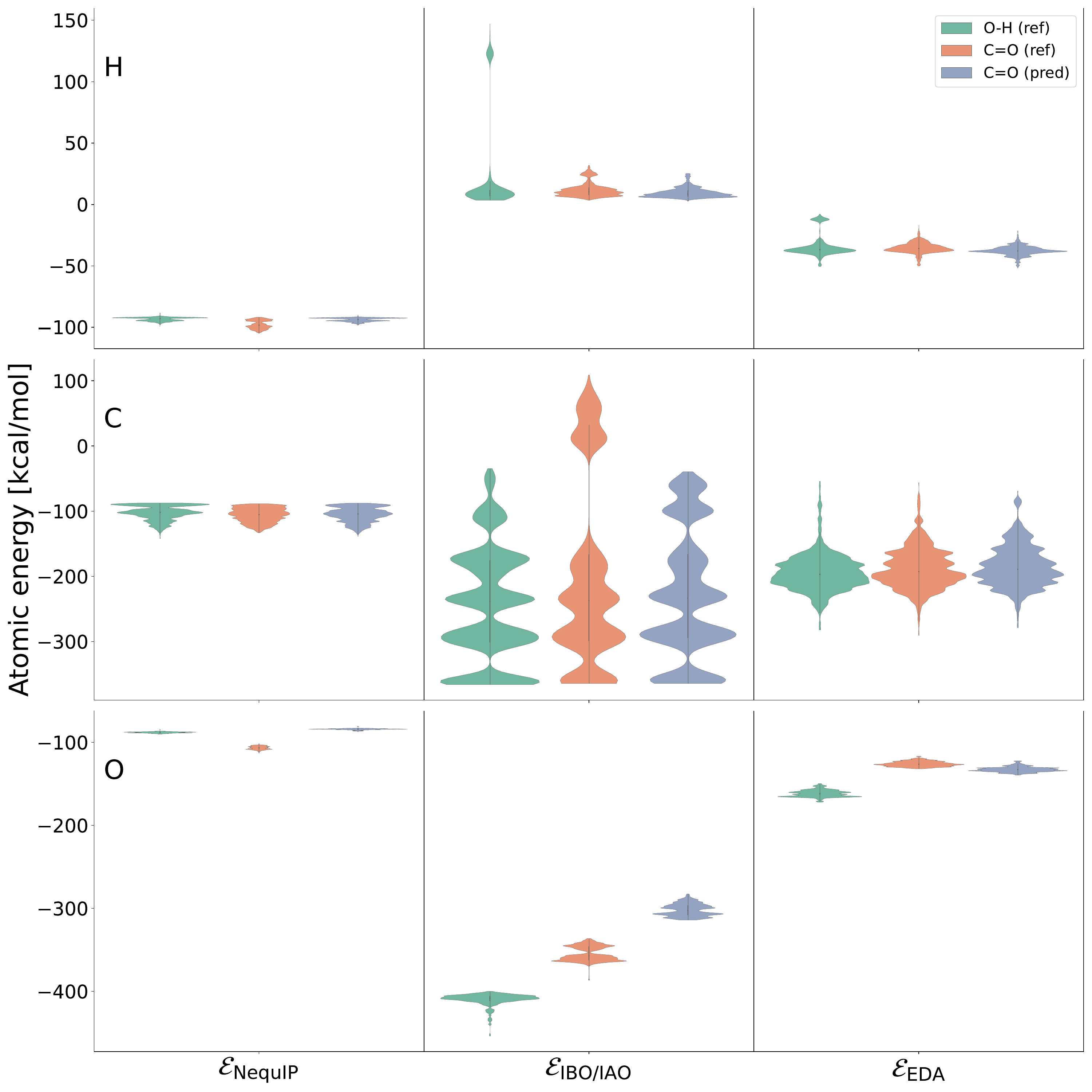}
    \caption{Distributions of B3LYP/pcseg-1 reference (ref) atomic energies across the datasets of hydroxyl (O--H) and carbonyl (C=O) compounds alongside predicted (pred) atomic energies for the latter. The energies obtained from the data-driven and two electronic-structure decompositions are denoted as $\mathcal{E}_{\mathrm{NequIP}}$, $\mathcal{E}_{\mathrm{IBO/IAO}}$, and $\mathcal{E}_{\mathrm{EDA}}$, respectively.}
    \label{fig:ref_ea_o}
\end{figure}

Fig. \ref{fig:ref_ea_o} reports distributions of reference and predicted atomic energies for the hydroxyl and carbonyl datasets obtained using either data-driven or electronic-structure decompositions (both with respect to energies of atoms in vacuum). In the case of a data-driven decomposition ($\mathcal{E}_{\mathrm{NequIP}}$), reference atomic energies are obtained by training and evaluating a standard {\texttt{NequIP}} model on the same dataset (based on total energies), which should then yield close to optimal atomic energies across this. Similar distributions of atomic energies for the primary and secondary amines are presented in Fig. S2 of the supporting information (SI).\\

As was previously studied in Ref. \citenum{kjeldal2023decomposing}, atomic energies from an AO-based partitioning scheme (EDA) in a modest-sized basis set without augmentation by diffuse functions (pcseg-1) tend to all be negative in value and cluster within rather narrow bands without much binning of contributions within these. This observation is confirmed in the results of Fig. \ref{fig:ref_ea_o}. In addition, the exact same pattern is observed to hold true, to an even greater extent, for the atomic energies returned by the standard {\texttt{NequIP}} model trained on total energies only. In fact, the separation between aliphatic and hydroxyl hydrogens is even less pronounced in the data-driven results, as is that between the different oxygens, for which the order of stabilization is even reversed with respect to both of the AO- and MO-based partitionings.\\

In the results of the MO-based decomposition (IBO/IAO) in Fig. \ref{fig:ref_ea_o}, a much clearer distinction is observed between the atoms of the hydroxyl and carbonyl compounds, in support of the earlier observations made in Ref. \citenum{kjeldal2023decomposing}. The same is true with respect to the separation of the nitrogens of the primary and secondary amines in Fig. S2, which also shows the nitrogens (and carbons) of heterocyclic amines as outliers in the energies of the latter set. In Fig. \ref{fig:ref_ea_o}, the hydroxyl hydrogens are well separated from those bonded to carbon atoms, as are those adjacent to the C=O groups, and different classes of carbon atoms are binned into individual bands. Among the oxygens of the carbonyl compounds, a distinction between aldehydes and ketones is even observed. Unlike the energies of both the standard {\texttt{NequIP}} model and the AO-based EDA partitioning, those of the MO-based IBO/IAO analogue thus clearly reflect differences in local chemical environments and the electronic structures these give rise to. In Fig. S1 of the SI, we have further isolated atomic energies of carbon atoms belonging to different functional groups to emphasize how the IBO/IAO partitioning is the only among the three in Fig. \ref{fig:ref_ea_o} that convincingly and effectively account for this distinction.\\

\begin{figure}[ht!]
    \centering
    \includegraphics[width=0.8\textwidth]{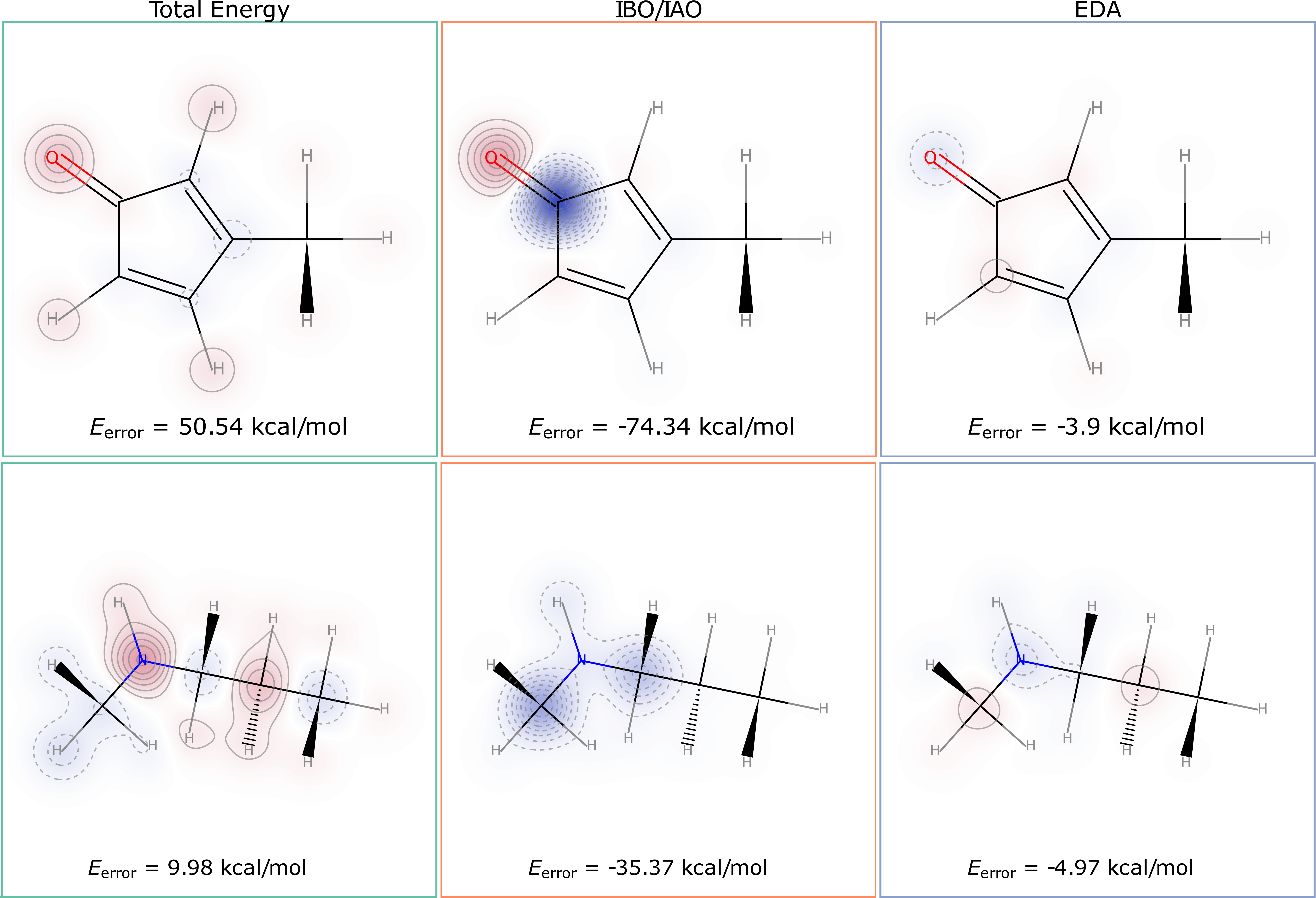}
    \caption{Errors in atomic energies for a random ketone (upper) and secondary amine (lower). Red and blue contours denote positive and negative errors, respectively. Errors have been normalized across the plots of the upper and lower panels to accentuate differences in results.}
    \label{fig:contour_o}
\end{figure}
From the reference distributions in Fig. \ref{fig:ref_ea_o}, alongside our prior knowledge of the hydroxyl and carbonyl datasets, one would expect an ML model trained exclusively on the former and evaluated on the latter to yield large errors, particularly given the unseen chemistry of the C=O groups. Fig. \ref{fig:contour_o} confirms this assumption for a random ketone, in that errors in energies associated with atoms in or close to the carbonyl groups are observed for all three models. From the predicted energies in Fig. \ref{fig:ref_ea_o}, the carbon energies in both the standard {\texttt{NequIP}} and EDA-based models are near mirror images of the reference values, while for the MO-based IBO/IAO model, the full destabilization of carbonyl carbons fails to manifest. In the same way, the oxygen energies change only marginally in the model based on EDA, which, given how the reference energies in-between the hydroxyl and carbonyl datasets practically (and fortuitously) coincide, lead to very low errors in the predictions of the trained model. The {\texttt{NequIP}} model, on the other hand, yields positive errors in the oxygen energies, as does the IBO/IAO model. In the former case, this happens because the oxygens are all predicted to be practically identical to the ones of the hydroxyl compounds, whereas in the latter case, all oxygen energies are subject to an upwards shift with respect to the reference values for the hydroxyl compounds, which, although correct, is ultimately too large in magnitude.\\

In terms of transferability across different classes of amines, this is observed to be a fundamentally more manageable task, cf. Fig. S2 of the SI. Errors are generally smaller than those observed for the application in-between hydroxyls and carbonyls, and only the standard (total energy) model exhibits basic problems in predicting the energies of the central nitrogen atoms. Statistics in support of Figs. \ref{fig:ref_ea_o}, \ref{fig:contour_o}, and S2 are presented in Tables S1 and S2 of the SI. Important in the context of Sect. \ref{adv_val_res_sect} and the discussions to follow in Sect. \ref{comp_trans_res_sect} is the fact that IBO/IAO-based errors tend to be large on or near unknown functional moieties, but the smallest among all three models upon moving away from these bond-by-bond.

\subsection{Compositional Transferability}\label{comp_trans_res_sect} 

We will now shift focus from functional to compositional transferability. Less so than HDNNs built around fixed descriptors, e.g., atom-centered symmetry functions, message-passing models with their learnable descriptors will still fundamentally rely on a set of atomic basis functions. For this reason alone, it is fair to assume some degree of equivalence between the data-driven decomposition of atomic energies from a model like {\texttt{NeuqIP}} and the energies yielded by an AO-based partitioning scheme like EDA, in which all energetic trace operations are restricted to the Gaussian basis functions spatially local to the individual atoms of a molecule. However, this will hold true only in basis sets without diffuse functions.\\

As such, when comparing distributions of atomic energies yielded by either {\texttt{NeuqIP}} or EDA, while not as similar as was reported for the fixed-descriptor results in Ref. \citenum{kjeldal2023decomposing}, the reference results across the QM7/13$^*$ and QM9/17$^*$ datasets in Figs. S3 and S4 of the SI largely show exactly this. Both sets of results are thus largely predetermined, as also evident from the fact that distributions of reference and predicted atomic energies are practically indistinguishable for both decompositions, regardless of which of the datasets one opts for, and results are further near-identical across all four of these. That being said, some differences between the {\texttt{NeuqIP}} and EDA reference results do exist, namely, in the noteworthy case of sulfur (Fig. S3), but on the whole the two resemble one another to a great extent. The results of the MO-based IBO/IAO decomposition, on the other hand, show much more diverse and structured distributions of atomic energies, with clear bands corresponding to specific atomic environments. Be that as it may, this increase in diversity among the atomic energies of the IBO/IAO decomposition, but also notable differences in the distributions of reference energies between training and testing datasets, are both perfectly reproduced in the results for QM13$^*$/17$^*$ returned by the models built around the IBO/IAO decomposition.\\

\begin{table}[ht!]
    \centering
    \begin{tabular}{c|c|c|c|c|c|c}
          \multirow{2}{*}{$l_{\mathrm{max}}$} & \multicolumn{2}{c|}{Total Energy} & \multicolumn{2}{c|}{IBO/IAO} & \multicolumn{2}{c}{EDA}\\
         \cline{2-7}
         & QM7/13$^*$ & QM9/17$^*$ & QM7/13$^*$ & QM9/17$^*$ & QM7/13$^*$ & QM9/17$^*$ \\
         \hline\hline
         $0$
         & 17.02 $\pm$ 1.41 & 6.06 $\pm$ 1.57 & 17.42 $\pm$ 1.82 & 5.25 $\pm$ 1.41 & 11.37 $\pm$ 1.38 & 7.48 $\pm$ 4.13 \\
         \hline
         $1$
         & 6.22 $\pm$ 0.33 & 1.79 $\pm$ 0.06 & 6.71 $\pm$ 0.32 & 2.46 $\pm$ 0.15 & 6.20 $\pm$ 0.34 & 3.36 $\pm$ 0.33 \\
         \hline
         $2$
         & 5.70 $\pm$ 0.53 & 1.36 $\pm$ 0.11 & 5.10 $\pm$ 0.05 & 1.77 $\pm$ 0.08 & 4.60 $\pm$ 0.11 & 2.28 $\pm$ 0.04
    \end{tabular}
    \caption{Molecular mean absolute errors (in kcal/mol) for the models trained on the QM7 and QM9 datasets and evaluated on QM13$^*$ and QM17$^*$, respectively. The mean and standard deviations are obtained through five independent training runs, except for $l_{\mathrm{max}} = 1, 2$ in the case of the model trained on QM9, for which only three independent runs were performed.}
    \label{tab:qmx_to_qmy}
\end{table}
In Table \ref{tab:qmx_to_qmy}, we report mean absolute errors (MAEs) for the transferability tests of the models trained on either QM7 or QM9 and applied to QM13$^*$ or QM17$^*$, respectively. We report three different model architectures, where the $l_{\mathrm{max}}$ parameter---corresponding to the highest rotation order allowed in the internal {\texttt{NeuqIP}} representation---is varied from 0 to 2. $l_{\mathrm{max}}=0$ thus provides no equivariance, corresponding to an invariant MPNN, while models with higher values of the $l_{\mathrm{max}}$ parameter encode more angular information into the models, at the expense of increased computational costs involved in the training phase. Training curves for $l_{\mathrm{max}} = 0-2$ are provided for the models trained on QM9 in Figs. S5--S7 of the SI.\\

For both transferability tests, we see a systematic decrease in the associated errors as $l_{\mathrm{max}}$ is increased for all three differently trained models and across both training sets (QM7/9). On the whole, the results in Table \ref{tab:qmx_to_qmy} appear to show that the individual models perform comparatively well, but also that the QM7 dataset is likely too limited in both size and composition (functional diversity, distinct motifs, etc.) to act as a realistic training pool for inferring electronic structures of larger molecular systems. Drawing also on our adversarial validation in Sect. \ref{adv_val_res_sect}, we will now proceed to inspect $(i)$ whether QM7 is indeed unfit for purpose and $(ii)$ if the different models yield comparative performances for the same reasons.\\

\begin{figure}[htbp]
    \centering
    \includegraphics[width=0.8\textwidth]{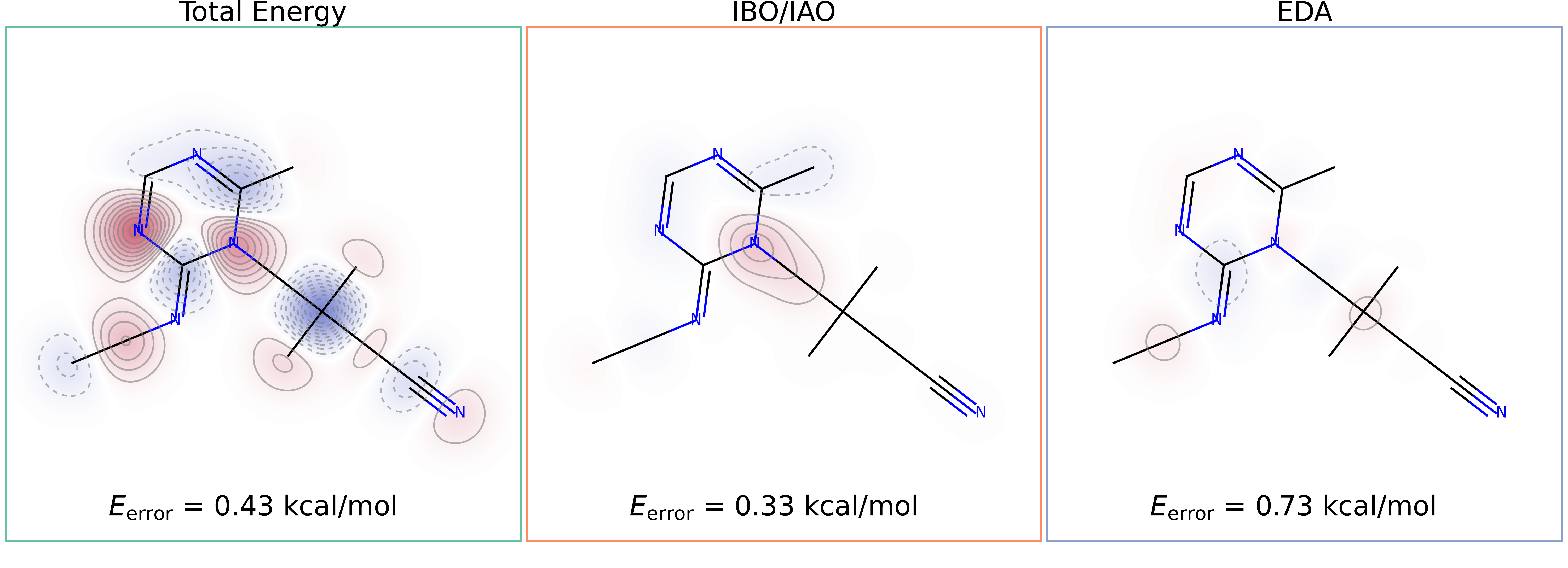}
    \caption{Errors in atomic energies for a random QM17$^*$ molecule ($l_{\mathrm{max}} = 1$).}
    \label{fig:qm17_contour}
\end{figure}
On par with Fig. \ref{fig:contour_o}, Fig. \ref{fig:qm17_contour} shows errors in predicted atomic energies of the different models trained on QM9 for a random molecule of the QM17$^*$ test set, with additional examples provided in Fig. S9 of the SI~\bibnote{Unlike in Fig. \ref{fig:contour_o}, potential errors in the atomic energies of hydrogens have been folded in onto the nearest heavy atom in the results of both Figs. \ref{fig:qm17_contour} and S9}. While only individual, selected examples, these results for the models based on IBO/IAO and EDA data appear to support the conclusions drawn from the earlier adversarial validation, namely, that the central atoms exhibit the largest errors. In contrast, for the model based exclusively on total energies across the QM9 dataset, the errors against a reference data-driven decomposition derived from QM17$^*$ itself are distributed across the entire molecule and seemingly lacking any systematic trends ({\textit{vide infra}}).\\

To further support these claims, we compute dataset-wide statistics for the individual atomic errors by using {\texttt{RDKit}} to identify central atoms and {\texttt{NetworkX}} to traverse outwards away from these in the graphs, one bond at a time, akin to what was done in Tables S1 and S2 of the SI~\cite{rdkit_prog,networkx}. In doing so, we choose to fold in all errors associated with hydrogen atoms onto their nearest heavy atom, like in Fig. \ref{fig:qm17_contour}; given how errors for hydrogens are uniformly small in magnitude, including these individually would risk conflating the general picture.\\

\begin{figure}[ht!]
    \centering
    \includegraphics[width=0.9\textwidth]{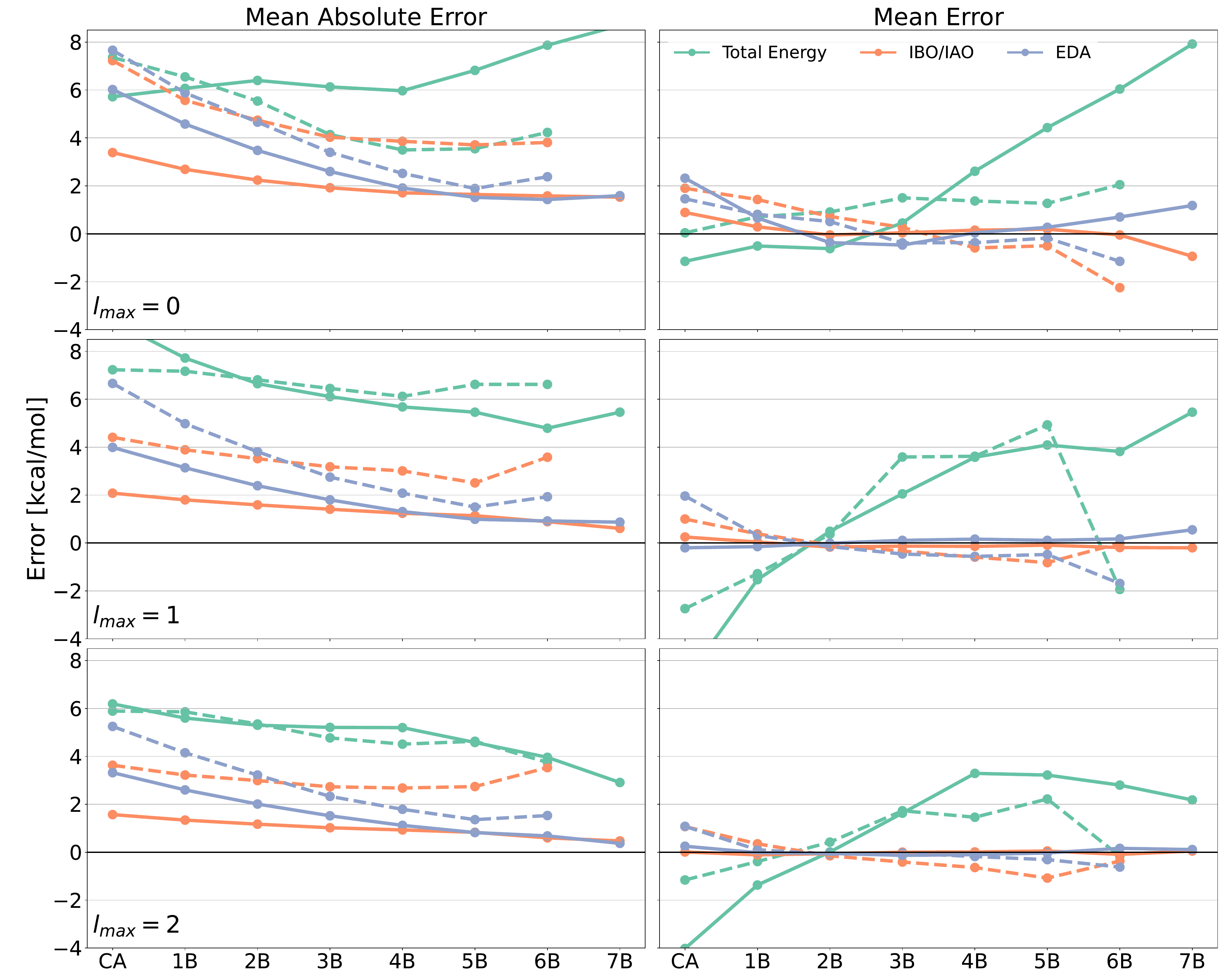}
    \caption{Mean absolute and signed errors (in kcal/mol) for the various models as we traverse $n$ bonds ($n$B) away from the central atoms (CA) in QM13$^*$ (dashed) and QM17$^*$ (solid).}
    \label{fig:qmx_neighbor_errors}
\end{figure}
The results in Fig. \ref{fig:qmx_neighbor_errors} collectively give rise to a number of key observations of great importance to the present study. First, the QM7 dataset is obviously not fit for the purpose of generalizing its chemistry to larger systems. As alluded to earlier, this is due to its very limited size but crucially also the fact that the diversity of its chemical composition is insufficient. A comparison of the learning curves in Fig. S8 of the SI further supports this observation, as does the alternative version of Fig. \ref{fig:qmx_neighbor_errors} in Fig. S10; here, a model built on a reduced QM9 training set of only 10k molecules (comparable to the size of QM7) still produces transferability results in moving from QM9 to QM17$^*$ that strongly resemble those in Fig. \ref{fig:qmx_neighbor_errors}. Regardless of the model of choice, and regardless of whether one chooses to gauge performance based on mean absolute or signed errors, the QM7-based curves in Fig. \ref{fig:qmx_neighbor_errors} are observed to plateau, with atomic errors in peripheral regions of the QM13$^*$ molecules often as large as those in the most central regions (which are otherwise expected to be the greatest).\\

Second, turning to the models trained exclusively on total energies, while these observe decent overall regression capabilities (particularly in the application to the QM17$^*$ dataset when trained on QM9, cf. Table \ref{tab:qmx_to_qmy} and Fig. S8), the atomic transferability in-between different training pools is inherently poor. Errors associated with the individual atoms are observed to be distinctively erratic and unsystematic, and optimal atomic energies thus differ significantly in moving from one training set to another, e.g., using either QM9 or QM17$^*$ for this purpose. In this context, it is important to note that large mean errors for these models in Fig. \ref{fig:qmx_neighbor_errors} do not necessarily translate into corresponding errors in predictions of total energies; that being said, the results in Fig. \ref{fig:qmx_neighbor_errors} succinctly show how atomic energies from data-driven compositions are but arbitrary variables and, thus, that these cannot reasonably be used to draw conclusions on local electronic structures and associated properties of these.\\

Finally, for the models based on the atomic energies of an MO-based IBO/IAO decomposition of QM9 (particularly for $l_{\mathrm{max}} \geq 1$), errors in ML predictions of the QM17$^*$ counterparts are not only observed to be small on average but also well spatially localized around the atoms of regions within the molecules of QM17$^*$ that are expected, on the grounds of adversarial validation, to be most foreign to a given model. While the models based on an AO-based EDA decomposition observe the same overall trend, they do so at significantly larger mean absolute errors for the innermost atoms. In addition, it is arguably worth reiterating once more how this type of decomposition is highly sensitive to the composition of an AO basis of choice, unlike the decomposition scheme based on the spatial locality of MOs instead. The IBO/IAO-based models are thus evidently the most systematic in the exercise concerned with inferring local electronic structures of the molecules in QM17$^*$ based on those present in QM9, a feature which ultimately lends itself to the fact that these local objects are both physically sound and unique in a robust MO-based decomposition like that based on a combination of IBOs and IAO weights. As such, the results in Fig. \ref{fig:qmx_neighbor_errors} hence give credence to the fundamental premise that local electronic structures around atoms embedded within extended molecules can indeed be successfully learned by means of contemporary, atom-based HDNNs, preferably ones that make proper use of equivariance rather than mere invariance.

\section{Discussion and Conclusions}\label{concl_sect}

In the present work, we have employed adversarial validation as a means to distinguish between different molecular datasets for the task of machine learning both total and atomic energies, with a view to analyzing and predicting when and how transferability between different chemically diverse datasets is to be expected and on what grounds. Using this technique, we have identified which specific parts (atomic regions) of a molecule any rigorous machine model is prone to exhibit larger errors for due to unseen chemical environments. We have demonstrated the usefulness of adversarial validation in two different contexts; first, through the application to two pathological, proof-of-concept examples, in which the datasets for training and testing were intentionally made to contain different functional groups. Second, adversarial validation was used to study transferability with respect to molecular composition by gauging when and how generalization to unseen chemistry will be sensible or not.\\ 

Our deduced similarities and possible discrepancies between any two datasets were next numerically tested by training equivariant neural networks on either total molecular energies only or a combination of these and a set of decomposed, atomic energies obtained at the level of Kohn-Sham density functional theory. In tests of both functional and compositional transferabilty---through applications between both pathological and realistic molecular datasets---we have found the inference of physically sound local electronic structures and properties to be feasible only whenever adequate knowledge of these is embedded into the training pool. In other words, only whenever an ML model is trained on sufficient information of the intrinsic local properties associated with different chemical functional groups and motifs will it be reasonable to expect the generalization of atomic errors to align with prior indications of the spatial parts of a molecule for which ML predictions should face difficulties.\\

We find the popular QM7 dataset to be too limited in both size and functional composition to warrant such generalizations to larger systems, while the more comprehensive QM9 dataset may indeed allow for this, given that the body of atomic energies available for training purposes satisfactorily exposes the uniqueness of different functional moieties within molecules. A decomposition scheme based on tailored spatially localized molecular orbitals (IBOs) and a set of appropriate atomic weights (determined from IAOs) has been shown to accommodate these requirements, while alternatives based on the spatial locality of atomic orbitals alone are deemed less fit due to being too insensitive to differences in local atomic environments (even disregarding the strong basis set dependence of such schemes).\\

Moving forward, we foresee that the use of befitting atomic energies for training high-dimensional neural networks will be beneficial, particularly in low-data regimes. More data points may be extracted from any given number of electronic-structure simulations and thus be made available in the training pool. Moreover, even for generalization purposes, where one desires to transfer performance in predictions from one diverse dataset to another, may such models have favourable advantages over the current standard of training only on total molecular energies. As we have demonstrated in the course of the present work, chemically intuitive atomic energies can indeed be inferred from common atom-based neural network architectures, especially whenever these implement equivariant features, and this paves the way towards being able to make qualified predictions of local energy contributions and bridge changes in these to key chemical concepts, such as, selectivity, reactivity, and stability.

\section*{Acknowledgments}

This work was supported by two research grants awarded to JJE, no. 37411 from VILLUM FONDEN (a part of THE VELUX FOUNDATIONS) and no. 10.46540/2064-00007B from the Independent Research Fund Denmark.

\section*{Supporting Information}

The supporting information (SI) contains two training configurations (as accompanying YAML files) for models based either exclusively on total energies or a combination of these and corresponding atomic energies, {\texttt{example\_total.yaml}} and {\texttt{example\_atom.yaml}}. In addition, Tables S1 and S2 report the spatial (bond-wise) distribution of atomic errors in relation to the results in Sect. \ref{func_trans_res_sect}, while Figs. S1 and S2 present further results on par with Fig. \ref{fig:ref_ea_o}, and Figs. S3 and S4 presents these same kind of results for the QM7/9 and QM13$^*$/17$^*$ datasets. Training curves for the models of Sect. \ref{comp_trans_res_sect} are provided in Figs. S5--S7, Fig. S8 presents corresponding learning curves, while Fig. S9 presents results similar to the ones in Fig. 5 but for two other entries of QM17$^*$. Finally, Fig. S10 presents a version of Fig. 6 for which the training set available to the QM9-based model was reduced to 10k molecules.

\section*{Data Availability}

Data in support of the findings of this study are available within the article, the supporting information, and in a dedicated Zenodo repository (DOI: 10.5281/zenodo.13837539).

%%%%%%%%%%%%%%%%%%%%%%%%%%%%%%%%%%%%%%%%%%

\providecommand{\latin}[1]{#1}
\makeatletter
\providecommand{\doi}
  {\begingroup\let\do\@makeother\dospecials
  \catcode`\{=1 \catcode`\}=2 \doi@aux}
\providecommand{\doi@aux}[1]{\endgroup\texttt{#1}}
\makeatother
\providecommand*\mcitethebibliography{\thebibliography}
\csname @ifundefined\endcsname{endmcitethebibliography}
  {\let\endmcitethebibliography\endthebibliography}{}


\begin{mcitethebibliography}{61}
\bibitem[Not()]{Note-1}
For the model configuration, we used the standard {\texttt{NequIP}} defaults.
  Example configuration files can be found among the SI as YAML files. In
  general, no optimization of hyperparameters was done to ensure a fair
  comparison between the different models.\relax
\mciteBstWouldAddEndPunctfalse
\mciteSetBstMidEndSepPunct{\mcitedefaultmidpunct}
{}{\mcitedefaultseppunct}\relax
\EndOfBibitem
\bibitem[Not()]{Note-3}
From the geometry optimizations, 3,814 molecules converged with no imaginary
  frequencies, of which 3,553 successfully passed a subsequent SCF stability
  check in {\texttt{PySCF}}.\relax
\mciteBstWouldAddEndPunctfalse
\mciteSetBstMidEndSepPunct{\mcitedefaultmidpunct}
{}{\mcitedefaultseppunct}\relax
\EndOfBibitem
\bibitem[Not()]{Note-4}
Unlike in Fig. \ref{fig:contour_o}, potential errors in the atomic energies of
  hydrogens have been folded in onto the nearest heavy atom in the results of
  both Figs. \ref{fig:qm17_contour} and S9\relax
\mciteBstWouldAddEndPuncttrue
\mciteSetBstMidEndSepPunct{\mcitedefaultmidpunct}
{\mcitedefaultendpunct}{\mcitedefaultseppunct}\relax
\EndOfBibitem
\end{mcitethebibliography}
\end{document}